%% file: main.tex
\documentclass[a4paper]{article}

\usepackage[english]{babel}
\usepackage[utf8x]{inputenc}
\usepackage[T1]{fontenc}

\usepackage[a4paper,top=2.54cm,bottom=2.54cm,left=2.54cm,right=2.54cm,marginparwidth=1.75cm]{geometry}

\usepackage{amsmath}
\usepackage{graphicx}
\usepackage[colorinlistoftodos]{todonotes}
\usepackage[colorlinks=false, allcolors=blue]{hyperref}
\usepackage{parskip}
\usepackage{setspace}

\usepackage{cite}

\usepackage{amsmath}

\usepackage{graphicx}

\usepackage{multirow}
\usepackage{float}
\usepackage{etoolbox}
\usepackage[font=large,labelfont=bf]{caption}
\usepackage{mwe}

\title{\textbf{Insight into the physical properties of two niobium based compounds Nb\textsubscript{3}Be and Nb\textsubscript{3}Be\textsubscript{2} via first principles calculation}}

\author{\normalsize Md. Zahidur Rahaman\textsuperscript{1}\\\vspace{.01in}
\\\small \textit{Department of Physics}\\\small \textit{Pabna University of Science and Technology, Pabna-6600, Bangladesh}\\\small \textit{Bangladesh University of Engineering and Technology, Dhaka, Bangladesh}\\\small \textit{\href{mailto:zahidur.physics@gmail.com}{zahidur.physics@gmail.com}}\\\vspace{.01in}\\
\normalsize Md. Lokman Ali\textsuperscript{2*}\\\vspace{.01in}
\\\small \textit{Department of Physics}\\\small \textit{Pabna University of Science and Technology, Pabna-6600, Bangladesh}\\\small \textit{\href{mailto:lokman.cu12@gmail.com}{lokman.cu12@gmail.com}}
\thanks{Corresponding author}
}

\date{\small (28 August, 2017)}

\begin{document}
\maketitle

\input{abstract}
\input{introduction}
\input{method}  
\input{result}

\input{conclusion}

\input{reference}
\input{figures}
\input{table}

\end{document}

%% file: abstract.tex
\begin{abstract}
\addcontentsline{toc}{section}{Abstract}
\normalsize
\singlespace{
We investigate the structural, electronic, mechanical and elastic
properties of two niobium based intermetallic compounds
Nb\textsubscript{3}Be and Nb\textsubscript{3}Be\textsubscript{2} by
using the DFT based theoretical method. A good agreement is found among
the structural parameters of both the phases with experimentally
evaluated parameters. For both the phases metallic conductivity is
observed while Nb\textsubscript{3}Be phase is more conducting than that
of Nb\textsubscript{3}Be\textsubscript{2} phase. Evaluated DOS at Fermi
level indicates that Nb\textsubscript{3}Be\textsubscript{2} phase is
electrically more stable than Nb\textsubscript{3}Be phase. For both
phases Nb-4d states is mostly responsible for metallic conductivity. The
study of total charge density and Mulliken atomic population reveal the
existence of covalent, metallic and ionic bonds in both intermetallics.
Both the phases are mechanically stable in nature while
Nb\textsubscript{3}Be phase is more ductile than
Nb\textsubscript{3}Be\textsubscript{2} phase. The study of Vickers
hardness exhibits that Nb\textsubscript{3}Be\textsubscript{2} phase is
harder than that of Nb\textsubscript{3}Be. Both compounds are
anisotropic in nature while Nb\textsubscript{3}Be phase possesses large
anisotropic characteristics than that of
Nb\textsubscript{3}Be\textsubscript{2} phase. The Debye temperature of
both the compounds are also calculated and discussed.

}

\end{abstract}

%% file: introduction.tex
\clearpage
\section*{I. Introduction}
\addcontentsline{toc}{section}{Introduction}
\large 
\doublespacing

A-15 structured materials have gained huge interest in the research
community of solid state physics as well as material science for more
than five decades due to their many attractive physical properties. Most
of the compound in this family exhibits superconductivity; some of them
show high superconducting critical temperature. Many of them possess
unusual elastic and electrical properties. Many compounds of A-15 family
possess good corrosion resistance, low density and high melting point.
Not only in superconductor industry but also these intermetallics have a
number of applications in chemical industries, aerospace industry,
aircraft, biomedical instrumentations and hydrogen storage systems.

In 1931 Hartman et al. first observed the cubic A-15 structure {[}1{]}.
A-15 phases possess the A\textsubscript{3}B type composition, where A =
any transition metal and B = element from right side of the periodic
table {[}2-4{]}. Among the A\textsubscript{3}B type compounds, much
interest has been drawn those intermetallics which shows
superconductivity. For example, vanadium gallium (V\textsubscript{3}Ga)
is often used in superconducting electromagnet {[}5{]}. The critical
temperature and upper critical field of V\textsubscript{3}Ga is 14.2 K
and 19 Tesla respectively {[}6{]}. However, niobium (Nb) constitutes a
number of superconductors with a wide range of
\emph{T\textsubscript{c}} , such as Nb\textsubscript{3}Sb (0.2 K),
Nb\textsubscript{3}In (9.2 K), Nb\textsubscript{3}Bi (3 K),
Nb\textsubscript{3}Al (18.8 K), Nb\textsubscript{3}Ge (23.6 K),
Nb\textsubscript{3}Sn (18.9 K), Nb\textsubscript{3}Ga (20.2 K) {[}7{]}.
All these compounds belong to Cr\textsubscript{3}Si~structure type
{[}8-14{]}. Superconductivity in niobium beryllide
(Nb\textsubscript{3}Be) was first observed by Tuleushev et al. in 2003
{[}15{]}. They used thermal treatment of the amorphous film system to
synthesize Nb\textsubscript{3}Be. They obtained X-ray structural data to
identify the structure of Nb\textsubscript{3}Be. They also determined
the superconducting critical temperature (\emph{T\textsubscript{c}} ) of
Nb\textsubscript{3}Be as 10.0 K. However, except the structural and
superconducting properties of Nb\textsubscript{3}Be there is no more
information available in literature. On the other hand,
Nb\textsubscript{3}Be\textsubscript{2} phase possesses the same reactant
element as Nb\textsubscript{3}Be phase though their chemical composition
is quite different. So it is very interesting to study and compare the
physical properties of these two different phases of Nb-Be system.
Nb\textsubscript{3}Be\textsubscript{2} phase was first synthesized by
Zalkin et al. in 1960 by using X-ray powder diffraction technique
{[}16{]}. In their study, they found that in
Nb\textsubscript{3}Be\textsubscript{2} phase the powder pattern
indicates the presence of a fcc unit cell with lattice constant 10.94 Å.
Though they concluded the selected phase as an impurity or an additional
phase in Nb-Be system. However, except the structural determination no
further investigation has been done on
Nb\textsubscript{3}Be\textsubscript{2} phase up to now.

Therefore, in this present work we aim to study the detailed physical
properties of Nb\textsubscript{3}Be and
Nb\textsubscript{3}Be\textsubscript{2} phases by theoretical means. A
thorough comparison among the obtained physical properties of these two
phases have also been represented and discussed from the theoretical
view point.

%% file: method.tex
\section*{II. Method of computation}
\addcontentsline{toc}{section}{Method of computation}
\large 
\doublespacing

All theoretical calculations in the present work were performed by using
UPPW (Ultrasoft Pseudopotential Plane Wave) method based on DFT (Density
Functional Theory) implemented within CASTEP (Cambridge Serial Total
Energy Package) code {[}17{]}. GGA (Generalized Gradient Approximation)
was used to describe the exchange-correlation functional parametrized by
Perdew-Burke-Ernzerhof (PBE) {[}18-21{]}. For pseudo atomic calculation
Nb-4s\textsuperscript{2} 4p\textsuperscript{6} 4d\textsuperscript{4}
5s\textsuperscript{1} and Be-2s\textsuperscript{2} for both the
compounds were taken as valence electron. For constructing the k-point
meshes for the sake of Brillouin zone sampling Monkhorst-Pack scheme
{[}22{]} was used. For converging the total energy 8×8×8 grids with 400
eV cutoff energy for Nb\textsubscript{3}Be and 8×8×8 grids with 350 eV
cutoff energy for Nb\textsubscript{3}Be\textsubscript{2} were set. The
full geometry optimization of both the compounds was performed within
BFGS (Brodyden-Fletcher-Goldfarb-Shanno) scheme {[}23{]}. The elastic
stiffness constants of Nb\textsubscript{3}Be and
Nb\textsubscript{3}Be\textsubscript{2} intermetallics were computed by
using the stress-strain method {[}24{]}. The maximum ionic displacement
was fixed to 2.0 × 10\textsuperscript{-4} Å.

%% file: result.tex
\section*{III. Results and discussion}
\addcontentsline{toc}{section}{Results and discussion}

\subsection*{A. Structural Properties}
\addcontentsline{toc}{subsection}{Structural Properties}
\large 
\doublespacing

A-15 structured superconductor Nb\textsubscript{3}Be possesses cubic
crystal structure with \emph{PM-3N} (223) space group {[}15{]} whereas
Nb\textsubscript{3}Be\textsubscript{2} phase possesses tetragonal
crystal structure with space group \emph{P4/MBM} (127) {[}16{]}.
Nb\textsubscript{3}Be has eight atoms per unit cell with two formula
units whereas Nb\textsubscript{3}Be\textsubscript{2} phase has ten atoms
per unit cell with two formula units. The detailed fractional
coordinates of Nb\textsubscript{3}Be\textsubscript{2} can be found
elsewhere {[}16{]}. The fully relaxed crystal structures of both the
compounds are illustrated in Fig. 1. The computed lattice constants
\emph{a\textsubscript{0}} and \emph{c\textsubscript{0}} , cell volume
\emph{V\textsubscript{0}} and bulk modulus \emph{B\textsubscript{0}} of
both the phases are listed in Table 1 along with the experimental
values. From Table 1 it can be noticed that the computed structural
parameters in this study are in good agreement with the experimentally
evaluated parameters. Theoretically evaluated lattice constant shows
minor deviation with experimental value bearing the reliability of this
present study.


\subsection*{B. Single and Polycrystalline Elastic Properties}
\addcontentsline{toc}{subsection}{Single and Polycrystalline Elastic Properties}
\large 
\doublespacing

The cubic and tetragonal solids have three (\emph{C\textsubscript{11}} ,
\emph{C\textsubscript{12}} and \emph{C\textsubscript{44}} ) and six
(\emph{C\textsubscript{11}} , \emph{C\textsubscript{12}} ,
\emph{C\textsubscript{13}} , \emph{C\textsubscript{33}} ,
\emph{C\textsubscript{44}} and \emph{C\textsubscript{66}} ) independent
elastic constants respectively. These elastic constants can be achieved
by computing the total energy as a function of strain {[}25{]}. The
detailed of these calculations are explained elsewhere {[}26, 27{]}. The
obtained elastic constants of both the phases are recorded in Table 2.
There is no previous data available in literature about the elastic
constants of these two phases. Hence, this present study will be a
valuable source of reference for future experimental work. As shown in
Table 2, the value of \emph{C\textsubscript{33}} is larger than that of
\emph{C\textsubscript{11}} for Nb\textsubscript{3}Be\textsubscript{2}
implying that the incompressibility toward {[}001{]} direction is
stronger than {[}100{]} direction {[}28{]}. We also notice that
\emph{C\textsubscript{44}} is smaller than \emph{C\textsubscript{66}} 
indicating that {[}100{]}(010) shear is harder than {[}100{]}(001) shear
for Nb\textsubscript{3}Be\textsubscript{2} {[}28{]}. It can be noted
that almost all the elastic constants of Nb\textsubscript{3}Be are
slightly smaller than that of Nb\textsubscript{3}Be\textsubscript{2}
phase implying the weaker shear resistance and incompressibility of
Nb\textsubscript{3}Be compared with
Nb\textsubscript{3}Be\textsubscript{2}. The calculated elastic constants of both the phases are also compared with Nb\textsubscript{3}Ga as shown in Table 2.

For being mechanically stable the strain energy of a crystal must be
positive for homogeneous elastic deformation of the crystal {[}28{]}.
The Born stability criteria for tetragonal solids are given below.

\emph{C\textsubscript{11}} \textgreater{} 0, \emph{C\textsubscript{33}} 
\textgreater{} 0, \emph{C\textsubscript{66}} \textgreater{} 0,
\emph{C\textsubscript{44}} \textgreater{} 0, \emph{C\textsubscript{11}} + \emph{C\textsubscript{33}} -- 2\emph{C\textsubscript{13}} \textgreater{} 0, \emph{C\textsubscript{11}} -- \emph{C\textsubscript{12}} \textgreater{} 0, 2(\emph{C\textsubscript{11}} + \emph{C\textsubscript{12}} ) +
\emph{C\textsubscript{33}} + 4\emph{C\textsubscript{13}} \textgreater{}
0\\
For cubic crystal there are only three elastic constants and hence the
stability criteria is written as,

\emph{C\textsubscript{11}} \textgreater{} 0, \emph{C\textsubscript{44}} 
\textgreater{} 0, \emph{C\textsubscript{11}} --
\emph{C\textsubscript{12}} \textgreater{} 0 and
\emph{C\textsubscript{11}} + 2\emph{C\textsubscript{12}} \textgreater{}
0\\
Evidently, the above two phases of Nb are mechanically stable in nature
as their computed elastic constant data satisfy the respective stability
criteria as shown in Table 2. The elastic moduli of polycrystalline
compounds can be achieved from the single crystal elastic constants data
by using VRH (Voigt-Reuss-Hill) scheme {[}29{]}. This scheme provides
reasonably satisfactory data of elastic constants which has been
validated experimentally. For tetragonal solids the bulk and shear
modulus in this approximation are given as follows:
\begin{equation}
\tag{1}
B_{V} = \ \frac{2C_{11} + \ 2C_{12} + C_{33} + 4C_{13}}{9}
\end{equation}
\begin{equation}
\tag{2}
B_{R} = \ \frac{C^{2}}{M}
\end{equation}
\begin{equation}
\tag{3}
G_{V} = \ \frac{M + 3C_{11} - 3C_{12} + \ 12C_{44} + \ 6C_{66}}{30}
\end{equation}
\begin{equation}
\tag{4}
G_{R} = \ \frac{15}{\left\lbrack \frac{18B_{V}}{C^{2}} + \ \frac{6}{(C_{11} - \ C_{12})} + \ \frac{6}{C_{44}} + \ \frac{3}{C_{66}} \right\rbrack}
\end{equation}
Where,
\begin{equation}
\tag*{}
M = \ C_{11} + C_{12} + 2C_{33} - 4C_{13}
\end{equation}
And 
\begin{equation}
\tag*{}
C^{2} = \left( C_{11} + C_{12} \right)C_{33} - \ 2C_{13}^{2}
\end{equation}
For cubic crystal we get,
\begin{equation}
\tag{5}
{\ B}_{v} = \ B_{R} = \ \frac{(C_{11} + \ 2C_{12})}{3}
\end{equation}
\begin{equation}
\tag{6}
G_{v} = \ \frac{(C_{11} - \ C_{12} + \ 3C_{44})}{5}
\end{equation}
\begin{equation}
\tag{7}
G_{R} = \ \frac{5C_{44}(C_{11} - C_{12})}{\lbrack 4C_{44} + 3\left( C_{11} - \ C_{12} \right)\rbrack}
\end{equation}
Now, the value of \emph{B} and \emph{G} can be obtained as follows,
\begin{equation}
\tag{8}
B = \ \frac{1}{2}\left( B_{R} + \ B_{v} \right)
\end{equation}
\begin{equation}
\tag{9}
G = \ \frac{1}{2}\left( G_{v} + \ G_{R} \right)
\end{equation}
The Poisson's ratio, \emph{\(\nu\)} and Young's modulus, \emph{E} can be
estimated by using the following formulas,
\begin{equation}
\tag{10}
\nu = \ \frac{3B - 2G}{2(3B + G)}
\end{equation}
\begin{equation}
\tag{11}
E = \ \frac{9\text{GB}}{3B + G}
\end{equation}
The estimated polycrystalline elastic moduli are tabulated in Table 3. The calculated elastic moduli of both the phases are also compared with Nb\textsubscript{3}Ga.
As shown in Table 3 the bulk modulus of
Nb\textsubscript{3}Be\textsubscript{2} phase is slightly larger than the
Nb\textsubscript{3}Be phase implying the stronger resistance to change
in volume of Nb\textsubscript{3}Be\textsubscript{2} phase under external
pressure {[}30{]}. The shear modulus of
Nb\textsubscript{3}Be\textsubscript{2} phase is comparatively large than
that of Nb\textsubscript{3}Be phase indicating strong shear resistance
and strong covalent bond in Nb\textsubscript{3}Be\textsubscript{2} phase
{[}31{]}. The ratio between axial strain and uniaxial stress is
generally defined as the Young's modulus, which provides information
about the stiffness of solids {[}32{]}. Comparatively large value of
Young's modulus of Nb\textsubscript{3}Be\textsubscript{2} phase
indicates that Nb\textsubscript{3}Be\textsubscript{2} is stiffer than
the Nb\textsubscript{3}Be phase.

In order to identify the intrinsic ductility of solids the Poisson's
ratio is a useful index. The higher the Poisson's ratio, the more
ductile the material is. According to the value of Poisson's ratio
\emph{\(\nu\)} (Table 3) both the phases are ductile and ductility of
Nb\textsubscript{3}Be phase is better than
Nb\textsubscript{3}Be\textsubscript{2} phase. The Poisson's ratio is
also used to predict the bonding force exist in a solid. As shown in
Table 3 both the phases exhibit central force characteristics as the
value of \emph{\(\nu\)} from 0.25 to 0.50 indicates the existence of central force in a solid {[}33{]}. Another empirical criterion to judge the brittleness and ductility of solids is Pugh ratio (the ratio between the bulk modulus \emph{B} and shear modulus \emph{G} , \emph{B} /\emph{G} )
{[}32{]}. As shown in Table 3 both the compounds exhibit ductile manner
as \emph{B} /\emph{G} \textgreater{} 1.75. Though, Nb\textsubscript{3}Be
phase is more ductile than Nb\textsubscript{3}Be\textsubscript{2} phase.
This result accords well with that predicted form the value of Poisson's
ratio. Cauchy pressure is another useful index to explain the angular
character of atomic bonding in solids {[}34{]}. For tetragonal phase the
Cauchy pressures (\emph{C\textsubscript{13}} --
\emph{C\textsubscript{44}} ) and (\emph{C\textsubscript{12}} --
\emph{C\textsubscript{66}} ) as well as for cubic phase the Cauchy
pressure (\emph{C\textsubscript{12}} -- \emph{C\textsubscript{44}} ) are
evaluated and listed in Table 3. Obviously the positive value of Cauchy
pressure for both the compounds demonstrates the ductile manner {[}35{]}
consistent with above prediction.

The Vickers hardness is a very popular index to get information about
the hardness of a material. In this study we have used a very simple
potential formula developed by Chen et al {[}36{]} for calculating the
hardness of Nb\textsubscript{3}Be and
Nb\textsubscript{3}Be\textsubscript{2} phases given as follows,
\begin{equation}
\tag{12}
H_{V} = 2\left( K^{2}G \right)^{0.585} - \ 3
\end{equation}
Where, K is defined as the ratio between shear modulus and bulk modulus.
Evaluated values of \emph{H\textsubscript{v}} using eq. 12 for both
phases are listed in Table 3. Obviously, the computed values represent
similar trend as predicted above by Young's modulus that
Nb\textsubscript{3}Be\textsubscript{2} phase is harder than that of
Nb\textsubscript{3}Be.

The anisotropic characteristics of a tetragonal crystal can be computed
by the following equation {[}37{]},
\begin{equation}
\tag{13}
A^{U} = \ \frac{5G_{V}}{G_{R}} + \ \frac{B_{V}}{B_{R}} - \ 6
\end{equation}
For cubic crystal the evaluated value of bulk modulus is same according
to Voigt and Reuss approximation and hence Eq. 13 can be simplified as,
\begin{equation}
\tag{14}
A^{U} = \ 5\left( \frac{G_{V}}{G_{R}}\  - \ 1\ \  \right)
\end{equation}
The evaluated values of \emph{A\textsuperscript{U}} by using Eq. 13 and
Eq. 14 for tetragonal Nb\textsubscript{3}Be\textsubscript{2} and cubic
Nb\textsubscript{3}Be phase are tabulated in Table 3. For completely
isotropic material \emph{A\textsuperscript{U}} = 0, and deviation from
zero indicates the degree of anisotropy. As shown in Table 3 both the
phases are anisotropic in nature and Nb\textsubscript{3}Be phase
possesses large anisotropic characteristics than that of
Nb\textsubscript{3}Be\textsubscript{2} phase.


\subsection*{C. Electronic Properties and Chemical Bonding}
\addcontentsline{toc}{subsection}{Electronic Properties and Chemical Bonding}
\large 
\doublespacing

The electronic properties of cubic Nb\textsubscript{3}Be and tetragonal
Nb\textsubscript{3}Be\textsubscript{2} phase have been investigated
through the calculation of band structure, partial and total density of
states and total electron (charge) density. The band structures of both
the Nb-phase are illustrated in Fig. 2. Evidently, at Fermi level the
valence and conduction bands are overlapped implying the metallic nature
of both the phases. Though the superconducting properties of
Nb\textsubscript{3}Be have been investigated previously, the
superconducting properties of Nb\textsubscript{3}Be\textsubscript{2}
phase are still unexplored. The metallic nature of
Nb\textsubscript{3}Be\textsubscript{2} phase implies that this phase may
also possess superconducting characteristics {[}38{]}.

The total and partial density of states of Nb\textsubscript{3}Be and
Nb\textsubscript{3}Be\textsubscript{2} phases are illustrated in Fig. 3.
From -8 eV to 0 eV (Fermi level) the prevalent feature of hybridization
is noticed for Nb-4d and Be-2s states for both the intermetallics with
some contribution of Nb-5s and Nb-4p states. Nb-4d states contribute the
most to constitute the valence band of both phases. In conduction band
most of the contribution comes from Nb-4d states. Though, some of the
contribution comes from Nb-4p states. For both the compounds
approximately null contribution is observed for Be-2s states in
conduction band. However the contribution of Nb-4d states is most at
Fermi level. Nb metal is responsible for the metallic nature of both the
phases. The computed density of states at Fermi level is 13.82 states
per eV per unit cell for Nb\textsubscript{3}Be phase and 3.81 states per
eV per unit cell for Nb\textsubscript{3}Be\textsubscript{2} phase.
Therefore, Nb\textsubscript{3}Be phase is more conducting than that of
Nb\textsubscript{3}Be\textsubscript{2} phase. For metallic system, the
electronic stability depends upon the value of DOS at Fermi level. Metal
having lower value of \emph{N} (\emph{E\textsubscript{F}} ) shows more
stability than those having higher value of
\emph{N} (\emph{E\textsubscript{F}} ) {[}42{]}. According to this
condition Nb\textsubscript{3}Be\textsubscript{2} phase is electrically
more stable than Nb\textsubscript{3}Be phase.

For analyzing the bonding characteristics of Nb\textsubscript{3}Be and
Nb\textsubscript{3}Be\textsubscript{2} phases Mulliken atomic
populations {[}39{]} have been calculated. The data of atomic
populations is very handful to understand the chemical bonding
characteristics of materials. A null value of bond population implies
the ionic nature of that particular bond whereas high value indicates
the increasing level of covalency {[}40{]}. The estimated bond
populations of both the phases are listed in Table 4. It is evident from
Table 4 that the bond populations of both the compounds are positive and
greater than zero implying the existence of covalent bonds in both
phases. For both the phases Be atom carries the negative charges
indicating the transfer of electron from Nb to Be atom. Transferring of
charge from one atom to another precisely indicates the existence of
ionic bonds in both compounds. The corresponding bond lengths of both
phases are also shown in Table 4.

For further understanding the bonding nature in Nb\textsubscript{3}Be
and Nb\textsubscript{3}Be\textsubscript{2} phases the total charge
(electron) density is calculated along (001) plane as illustrated in
Fig. 4. A scale is shown at the right side of both plots indicating the
intensity of charge density. For both the compounds there are no
overlapping of electron distribution appeared. This result represents
the existence of ionic bond in Nb\textsubscript{3}Be and
Nb\textsubscript{3}Be\textsubscript{2}. But ionic nature is the result
of metallic character {[}41{]} indicating the existence of metallic
bonds in both phases. Therefore we can conclude that all the covalent,
ionic and metallic bonds are formed in Nb\textsubscript{3}Be and
Nb\textsubscript{3}Be\textsubscript{2} compounds and contribute equally
for the stability of both the phases.


\subsection*{D. Debye Temperature}
\addcontentsline{toc}{subsection}{Debye Temperature}
\large 
\doublespacing

The Debye temperature is the temperature which is associated with the
highest normal mode of oscillation of a crystal {[}43{]}. It is
associated directly or indirectly with many significant thermal
characteristics of crystals for example specific heat, melting point,
thermal expansion etc. So it is reasonable to determine the Debye
temperature of Nb\textsubscript{3}Be and
Nb\textsubscript{3}Be\textsubscript{2} phases. However, there are
various procedures and estimations available for computing the value of
\emph{\(\Theta\)\textsubscript{D}} . In this present study we have employed the
computed elastic constants to evaluate the Debye temperature of
Nb\textsubscript{3}Be and Nb\textsubscript{3}Be\textsubscript{2}
intermetallics. The average wave velocity (\emph{V\textsubscript{m}} ) of a solid is formulated as,
\begin{equation}
\tag{15}
\begin{split}
v_{m} = \ \left\lbrack \frac{1}{3}\left( \frac{2}{{v_{t}}^{3}} + \frac{1}{{v_{l}}^{3}} \right) \right\rbrack^{- \frac{1}{3}}
\end{split}
\end{equation}
Where, \emph{V\textsubscript{l}} and \emph{V\textsubscript{t}} are
longitudinal and transverse wave velocity respectively can be obtained
as follows,
\begin{equation}
\tag{16}
\begin{split}
v_{l} = \left( \frac{3B + 4G}{3\rho} \right)^{\frac{1}{2}}
\end{split}
\end{equation}
And
\begin{equation}
\tag{17}
\begin{split}
v_{t} = \left( \frac{G}{\rho} \right)^{\frac{1}{2}}
\end{split}
\end{equation}
Now, the Debye temperature (\emph{\(\Theta\)\textsubscript{D}} ) can be computed
as follows {[}44{]},
\begin{equation}
\tag{18}
\begin{split}
\theta_{D} = \ \frac{h}{k_{B}}\left( \frac{3N}{4\pi V} \right)^{\frac{1}{3}}{\times~ v}_{m}
\end{split}
\end{equation}
Where, \emph{K\textsubscript{B}} is the Boltzmann constant and \emph{h} 
is the Planck constant. The computed values of
\emph{V\textsubscript{l}} , \emph{V\textsubscript{t}} ,
\emph{V\textsubscript{m}} and \emph{\(\Theta\)\textsubscript{D}} for
Nb\textsubscript{3}Be and Nb\textsubscript{3}Be\textsubscript{2} phases
are listed in Table 5. As shown in Table 5 the Debye temperature of
Nb\textsubscript{3}Be\textsubscript{2} phase is larger than that of
Nb\textsubscript{3}Be phase. These values imply that for a single normal
vibration of Nb\textsubscript{3}Be crystal the highest temperature can
be achieved as 368.36 K and for Nb\textsubscript{3}Be\textsubscript{2}
crystal the highest temperature can be achieved as 450.91 K.

%% file: conclusion.tex
\section*{IV. Conclusions}
\addcontentsline{toc}{section}{Conclusions}
\large 
\doublespacing

In summary, the detailed physical properties including structural,
mechanical and electronic properties of two Nb-based intermetallic
compounds Nb\textsubscript{3}Be and
Nb\textsubscript{3}Be\textsubscript{2} have been explored by using
theoretical means. A good agreement is found among the structural
parameters of both the phases with experimentally evaluated parameters.
For both the phases metallic conductivity is observed while
Nb\textsubscript{3}Be phase is more conducting than that of
Nb\textsubscript{3}Be\textsubscript{2} phase. Evaluated DOS at Fermi
level indicates that Nb\textsubscript{3}Be\textsubscript{2} phase is
electrically more stable than Nb\textsubscript{3}Be phase. For both
phases Nb-4d states is mostly responsible for metallic conductivity with
minor contribution of other constituent orbitals. The study of total
charge density and Mulliken atomic population reveal the existence of
covalent, metallic and ionic bonds in both intermetallics. The metallic
nature of Nb\textsubscript{3}Be\textsubscript{2} phase implies that this
phase may also possess superconducting characteristics. The study of
elastic constants reveals that both intermetallics are mechanically
stable in nature while Nb\textsubscript{3}Be phase is more ductile than
Nb\textsubscript{3}Be\textsubscript{2} phase. Hence fabrication of
Nb\textsubscript{3}Be will be easier than
Nb\textsubscript{3}Be\textsubscript{2}. The study of Vickers hardness
exhibits that Nb\textsubscript{3}Be\textsubscript{2} phase is harder
than that of Nb\textsubscript{3}Be. Both compounds are anisotropic in
nature while Nb\textsubscript{3}Be phase possesses large anisotropic
characteristics than that of Nb\textsubscript{3}Be\textsubscript{2}
phase. The Debye temperature of Nb\textsubscript{3}Be is calculated to
be 368.36 K and for Nb\textsubscript{3}Be\textsubscript{2} phase 450.91
K. These values imply that for a single normal vibration of
Nb\textsubscript{3}Be crystal the highest temperature can be achieved as
368.36 K and for Nb\textsubscript{3}Be\textsubscript{2} crystal the
highest temperature can be achieved as 450.91 K. We hope the predicted
physical properties of these two compounds will motivate for
technological application of these compounds and also inspire other
researcher to conduct detailed experimental research on these
interesting materials in future.

%% file: reference.tex
\begin{center}
\begin{tikzpicture}
\draw (-2,0) -- (2,0);
\filldraw [black,very thick] (0,0) (-1,0) -- (1,0);
\filldraw [black,ultra thick] (0,0) (-.5,0) -- (.5,0);
\end{tikzpicture}
\end{center}

\begin{enumerate}
\def\labelenumi{\arabic{enumi}.}

\item
  H. Hartman, F. Ebert, O. Bretschneider, Z. Anorg. Allg. Chem. 198
  (1931) 116.
\item
  M. D. Banus, T. B. Reed, H. C. Gatos, M. C. Lavine, J. A. Kafalas, J.
  Phys. Chem. Solids 23 (1962) 971.
\item
  D. H. Killpatrick, J. Phys. Chem. Solids 25 (1964) 1213.
\item
  Y. Tarutani, U. Kawabe, Mater. Res. Bull. 13 (1978) 469.
\item
  \url{https://en.wikipedia.org/wiki/Vanadium-gallium}
\item
  Decker, D. L. Laquer, H. L. (1969), Magnetization Studies on Superconducting Vanadium Gallium (PDF), doi:10.1063/1.1658081
\item
  Sundareswari, M., Swaminathan Ramasubramanian, and Mathrubutham
  Rajagopalan. "Elastic and thermodynamical properties of A15
  Nb\textsubscript{3}X (X= Al, Ga, In, Sn and Sb) compounds---First
  principles DFT study."~\emph{Solid State Communications} ~150.41
  (2010): 2057-2060.
\item
  E.A. Wood, V.B. Compton, B.T. Matthias, E. Corenzwit, Acta
  Crystallogr. 11(1958) 604.
\item
  P.A. Beck (Ed.), Electronic Structure and Alloy Chemistry of the
  Transition Elements, Interscience Publishers, New York, 1963.
\item
  N.V. Nevit, in: J.H. Westbrook (Ed.), Intermetallics Compounds, R.E.
  Krieger Publishing Co., Huntington, NY, 1977.
\item
  M.D. Banus, T.B. Reed, H.C. Gatos, M.C. Lavine, J.A. Kafalos, J. Phys.
  Chem. Solids 23 (1962) 971.
\item
  D.H. Killpatrick, J. Phys. Chem. Solids 25 (1964) 1213.
\item
  Y. Tarutani, U. Kawabe, Mater. Res. Bull. 13 (1978) 469.
\item
  R. Flujkiger, H. Kupfer, J.L. Jorda, J. Muller, IEEE Trans. Magn. 23
  (1987) 980.
\item
  Tuleushev, A. Zh, V. N. Volodin, and Yu Zh Tuleushev. "Novel
  superconducting niobium beryllide Nb\textsubscript{3}Be with A15
  structure."~\emph{JETP Letters} ~78.7 (2003): 440-442.
\item
  Zalkin, A., D. E. Sands, and O. H. Krikorian. "Crystal structure of
  Nb\textsubscript{3}Be\textsubscript{2}."~\emph{Acta
  Crystallographica} ~13.2 (1960): 160-160.
\item
  Materials Studio CASTEP manual\_Accelrys, 2010. Pp. 261--262.
\item
  S.J. Clark, M.D. Segall, C.J. Pickard, P.J. Hasnip, M.J. Probert, K.
  Refson, M.C. Payne, Z.Kristallogr. 220 (2005) 567--570.
\item
  P. Hohenberg, W. Kohn, Phys. Rev. 136 (1964) B864--B871.
\item
  J.P. Perdew, A. Ruzsinszky, G.I. Csonka, O.A. Vydrov, G.E. Scuseria,
  L.A. Constantin, X. Zhou, K. burke, Phys. Rev. Lett. 100 (2008)
  136406--136409.
\item
  J.P. Perdew, A. Ruzsinszky, G.I. Csonka, O.A. Vydrov, G.E. Scuseria,
  L.A. Constantin, X. Zhou, K. Burke, Phys. Rev. Lett. 100 (2008)
  136406.
\item
  H. J. Monkhorst and J. D. Pack, Phys. Rev. B 13, 5188 (1976).
\item
  B. G. Pfrommer, M. Cote, S. G. Louie, and M. L. Cohen, J. Comput.
  Phys. 131, 233 (1997).
\item
  J. Kang, E.C. Lee, K.J. Chang, Phys. Rev. B 68 (2003) 054106.
\item
  L. Fast, J.Wills, B. Johansson, O. Eriksson, Phys. Rev. B 51 (1995)
  17431.
\item
  O. Beckstein, J. Klepeis, G. Hart, O. Pankratov, Phys. Rev. B 63
  (2001).
\item
  B.Y. Tang, N. Wang, W.Y. Yu, X.Q. Zeng, W.J. Ding, Acta Mater. 56
  (2008) 3353-3357.
\item
  Wu, Dong-Hai, et al. "First-principles study of structural stability
  and elastic properties of MgPd\textsubscript{3} and its
  hydride."~\emph{Journal of Magnesium and Alloys} ~2.2 (2014): 165-174.
\item
  Z.J.Wu, E.J. Zhao, H.P. Xiang, X.F. Hao, X.J. Liu, J. Meng, Phys. Rev.
  B 76 (2007) 054115.
\item
  L. Fast, J.Wills, B. Johansson, O. Eriksson, Phys. Rev. B 51 (1995)
  17431.
\item
  Z. Sun, S. Li, R. Ahuja, J.M. Schneider, Solid State Commun. 129
  (2004) 589e592.
\item
  S. Pugh, Philos. Mag. 45 (1954) 823e843.
\item
  P. Ravindran, L. Fast, P. Korzhavyi, B. Johansson, J. Wills, O.
  Eriksson, J. Appl. Phys. 84 (1998) 4891.
\item
  D. Pettifor, Mater. Sci. Technol. 8 (1992) 345e349.
\item
  D. Suetin, I. Shein, A. Ivanovskii, Solid State Sci. 12 (2010)
  814e817.
\item
  X.Q. Chen, H. Niu, D. Li and Y. Li, Intermetallics 19 (2011) p.1275.
\item
  S. I. Ranganathan and M. Ostoja-Starzewski, Phys. Rev. Lett. 101,
  055504 (2008).
\item
  Rahaman, Md Zahidur, and Md Atikur Rahman. "Novel Laves phase
  superconductor NbBe\textsubscript{2}: A theoretical
  investigation."~\emph{Computational Condensed Matter} ~8 (2016): 7-13.
\item
  R.S. Mulliken, J. Chem. Phys. 23 (1955) p.1833.
\item
  Segall, M. D.; Shah, R.; Pickard, C. J.; Payne, M. C. \emph{Phys. Rev.
  B} , 54, 16317-16320 (1996).
\item
  R.P. Singh, \emph{Journal of Magnesium and Alloys} 2 (2014) 349-356.
\item
  M.X. Zeng, R.N. Wang, B.Y. Tang, L.M. Peng, W.J. Ding, Model. Simul.
  Mater. Sci. Eng. 20 (2012) 035018.
\item
  Rahaman, Md Zahidur, and Md Atikur Rahman.
  "ThCr\textsubscript{2}Si\textsubscript{2}-type Ru-based
  superconductors LaRu\textsubscript{2}M\textsubscript{2} (M= P and As):
  An ab-initio investigation."~\emph{Journal of Alloys and
  Compounds} ~695 (2017): 2827-2834.
\item
  S. Aydin, M. Simsek, Phys. Rev. B: Condens. Matter 80 (2009) 134107.
\item
  Tian, Wenyan, and Haichuan Chen. "Theoretical investigation of the mechanical and thermodynamics properties of Nb\textsubscript{3}Ga superconductor under pressure." Journal of Alloys and Compounds 648 (2015): 229-236.
\item
  Sundareswari, M., Swaminathan Ramasubramanian, and Mathrubutham Rajagopalan. "Elastic and thermodynamical properties of A15 Nb\textsubscript{3}X (X= Al, Ga, In, Sn and Sb) compounds—First principles DFT study." Solid State Communications 150.41-42 (2010): 2057-2060.

\end{enumerate}

%% file: figures.tex
\clearpage
\begin{figure}[H]
\centering
\includegraphics[width=1\textwidth]{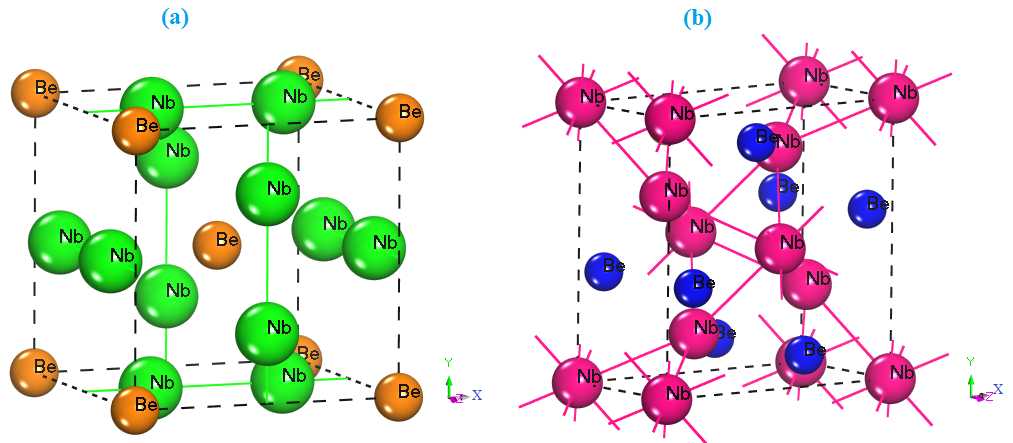}
\caption{The crystal structure of (a) Nb\textsubscript{3}Be and (b)
Nb\textsubscript{3}Be\textsubscript{2}.} 
\label{fig:Fig. 1}
\end{figure}

\begin{figure}[H]
\centering
\includegraphics[width=1\textwidth]{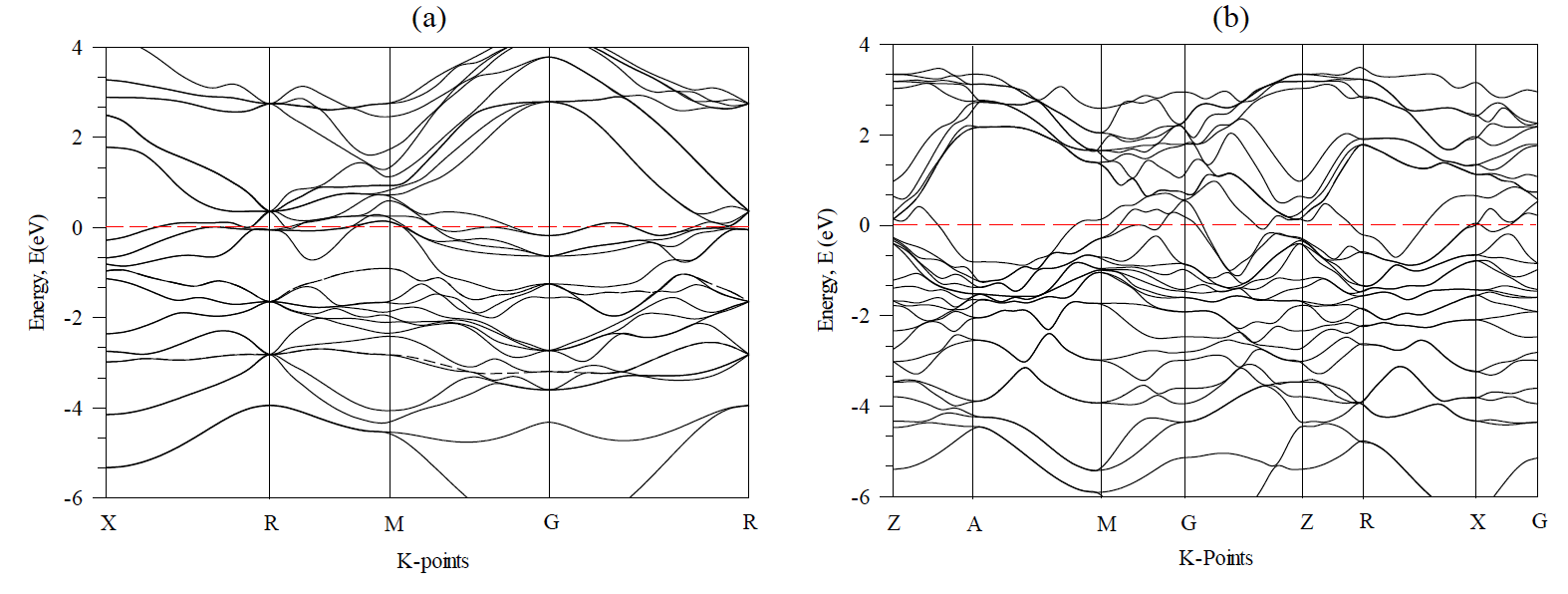}
\caption{The band structure of (a) Nb\textsubscript{3}Be and (b)
Nb\textsubscript{3}Be\textsubscript{2} phase.}  
\label{fig:Fig. 2}
\end{figure}

\begin{figure}[H]
\centering
\includegraphics[width=1\textwidth]{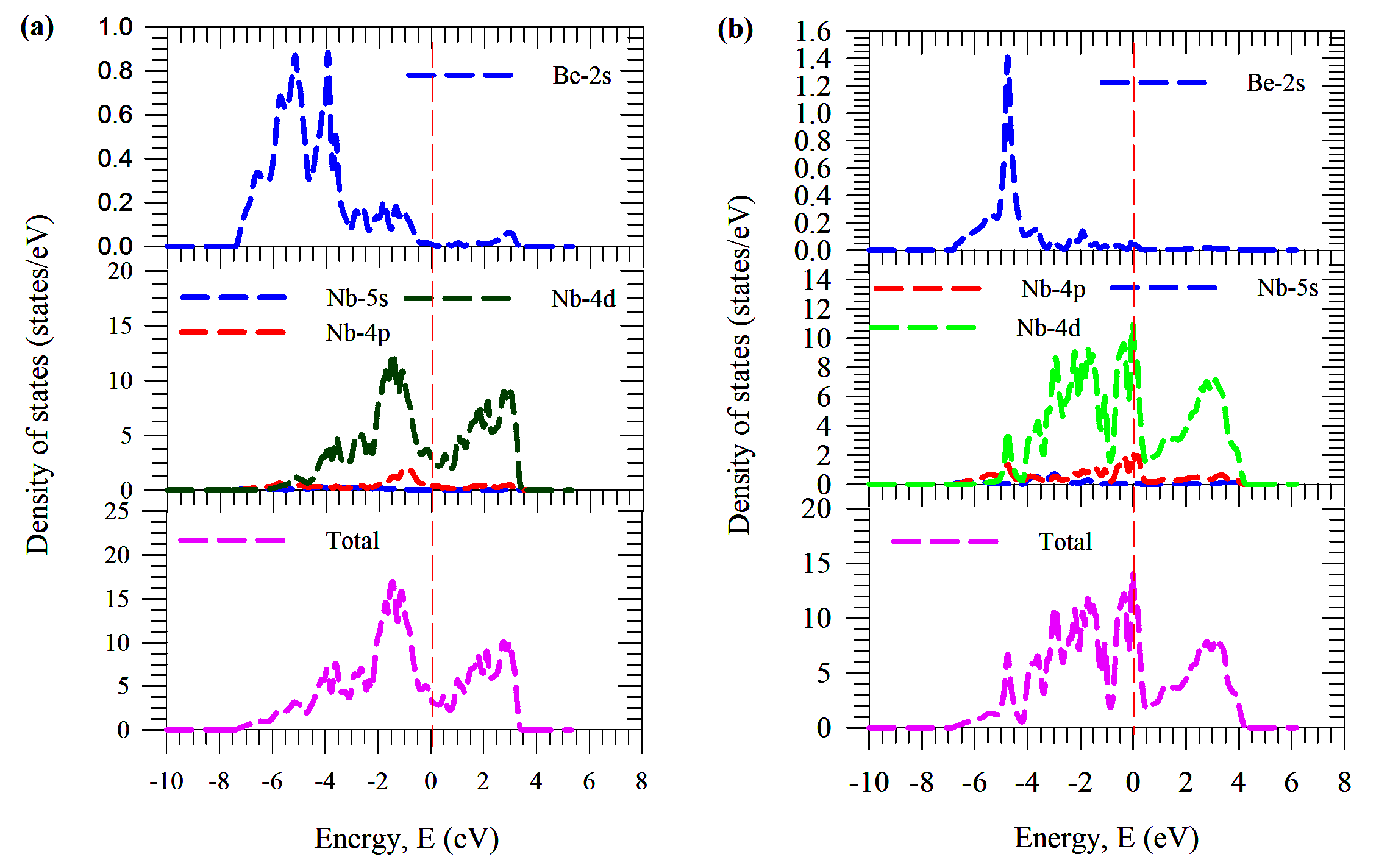}
\caption{Total and partial DOS of (a) Nb\textsubscript{3}Be\textsubscript{2}  and (b) Nb\textsubscript{3}Be intermetallics.}
\label{fig:Fig. 3}
\end{figure}

\begin{figure}[H]
\centering
\includegraphics[width=1\textwidth]{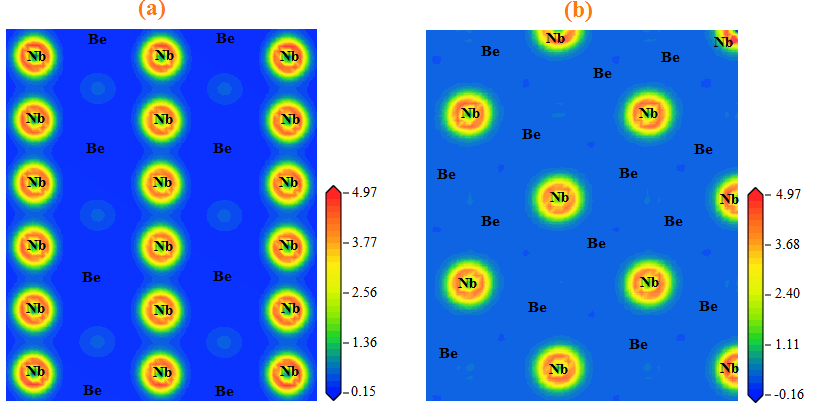}
\caption{Total charge density on (001) plane of (a) Nb\textsubscript{3}Be
and (b) Nb\textsubscript{3}Be\textsubscript{2} intermetallics.}
\label{fig:Fig. 4}
\end{figure}

%% file: table.tex
\clearpage

\begin{table*}[ht]
\doublespacing
\large
\centering 

\caption{\large Unit cell parameters of Nb\textsubscript{3}Be and
Nb\textsubscript{3}Be\textsubscript{2} intermetallics.}
\label{table 1}
\bigskip

\begin{tabular}[]{@{}cccccc@{}}
\hline
Properties & Nb\textsubscript{3}Be &&&
Nb\textsubscript{3}Be\textsubscript{2} \tabularnewline
\hline
&This study & Expt. {[}15{]} & & This study & Expt.
{[}16{]}\tabularnewline
\hline
\emph{a\textsubscript{0}} (Å ) & 5.070 & 5.187 & & 6.533 &
6.490\tabularnewline
\emph{c\textsubscript{0}} (Å ) & - & - & & 3.364 & 3.350\tabularnewline
\emph{c\textsubscript{0}/a\textsubscript{0}} & - & - & & 0.514 &
0.516\tabularnewline
\emph{V\textsubscript{0}} (Å\textsuperscript{3}) & 130.32 & 139.55 & &
143.57 & 141.10\tabularnewline
\emph{B\textsubscript{0}} (GPa) & 155.42 & - & & 162.23 &
-\tabularnewline
\hline
\end{tabular}

\end{table*}

\begin{table*}[ht]
\doublespacing
\large
\centering 

\caption{\large The computed elastic constants \emph{C\textsubscript{ij}} (in GPa) of Nb\textsubscript{3}Be and Nb\textsubscript{3}Be\textsubscript{2} phases.}
\label{table 2}
\bigskip

\begin{tabular}[]{@{}ccccccc@{}}
\hline
Compounds & \emph{C\textsubscript{11}} & \emph{C\textsubscript{12}} &
\emph{C\textsubscript{13}} & \emph{C\textsubscript{33}} &
\emph{C\textsubscript{44}} & \emph{C\textsubscript{66}}\tabularnewline
\hline
Nb\textsubscript{3}Be & 275.17 & 89.92 & - & - & 40.99 &
-\tabularnewline
Nb\textsubscript{3}Be\textsubscript{2} & 264.46 & 91.07 & 108.62 &
250.13 & 69.88 & 81.62\tabularnewline
Nb\textsubscript{3}Ga & 245.50\textsuperscript{a}, 305.41\textsuperscript{b} & 123.60\textsuperscript{a}, 104.90\textsuperscript{b} & - & - & 39.50\textsuperscript{a}, 48.78\textsuperscript{b} &
-\tabularnewline
\hline
\end{tabular}
\textsuperscript{a}Ref. 45; \textsuperscript{b}Ref. 46
\end{table*}

\begin{table*}[ht]
\doublespacing
\small
\centering 

\caption{\large Computed bulk modulus \emph{B} (GPa), shear modulus \emph{G} (GPa), Young's modulus \emph{E} (GPa), \emph{B/G} values, Poisson's ratio \emph{\(\nu\)} , Cauchy pressures (\emph{C\textsubscript{12}} --\emph{C\textsubscript{66}} ), (\emph{C\textsubscript{13}} --
\emph{C\textsubscript{44}} ) and (\emph{C\textsubscript{12}} --
\emph{C\textsubscript{44}} ) elastic anisotropy
\emph{A\textsuperscript{U}} and Vickers hardness
\emph{H\textsubscript{v}} (GPa) of Nb\textsubscript{3}Be and
Nb\textsubscript{3}Be\textsubscript{2} phases.}
\label{table 3}
\bigskip

\begin{tabular}[]{@{}ccccccccccc@{}}
\hline
Compounds & \emph{B} & \emph{G} & \emph{E} & \emph{B/G} & \emph{\(\nu\)} &
(\emph{C\textsubscript{12}} -- \emph{C\textsubscript{66}}) &
(\emph{C\textsubscript{13}} -- \emph{C\textsubscript{44}}) &
(\emph{C\textsubscript{12}} -- \emph{C\textsubscript{44}}) &
\emph{A\textsuperscript{U}} & \emph{H\textsubscript{v}}\tabularnewline
\hline
Nb\textsubscript{3}Be & 151.67 & 57.19 & 152.41 & 2.65 & 0.33 & - & - &
48.93 & 0.84 & 3.81\tabularnewline
Nb\textsubscript{3}Be\textsubscript{2} & 155.07 & 74.59 & 192.84 & 2.07
& 0.29 & 9.45 & 38.74 & - & 0.14 & 7.58\tabularnewline
Nb\textsubscript{3}Ga & 164.30\textsuperscript{a} & 47.0\textsuperscript{a} & 128.80\textsuperscript{a} & 3.49\textsuperscript{a} & 0.36\textsuperscript{a} & - & - &
- & 0.22\textsuperscript{a} & 3.30\textsuperscript{a}\tabularnewline
 & 169.97\textsuperscript{b} & 65.49\textsuperscript{b} & 174.11\textsuperscript{b} & - & 0.30\textsuperscript{b} & - & - &
55.31\textsuperscript{b} & - & -\tabularnewline
\hline
\end{tabular}
\textsuperscript{a}Ref. 45; \textsuperscript{b}Ref. 46
\end{table*}

\begin{table*}[ht]
\doublespacing
\large
\centering 

\caption{\large Mulliken atomic populations of cubic Nb\textsubscript{3}Be and tetragonal Nb\textsubscript{3}Be\textsubscript{2} phase.}
\label{table 4}
\bigskip

\begin{tabular}[]{@{}cccccccccc@{}}
\hline
Compounds & Species & s & p & d & Total & Charge & Bond & Population &
Length (Å)\tabularnewline
\hline
Nb\textsubscript{3}Be & Nb & 2.19 & 6.60 & 4.13 & 12.92 & 0.08 & Nb-Nb &
0.48 & 2.5350\tabularnewline
& Be & 0.54 & 1.69 & 0.00 & 2.23 & - 0.23 & Be-Nb & 0.37 &
2.8343\tabularnewline
Nb\textsubscript{3}Be\textsubscript{2 } & Nb & 2.29 & 6.66 & 4.03 &
12.97 & 0.03 & Be-Be & 0.44 & 2.0806\tabularnewline
& Be & 0.56 & 1.90 & 0.00 & 2.46 & - 0.46 & Be-Nb(1) & 0.56 &
2.5447\tabularnewline
& & & & & & & Be-Nb(2) & 0.34 & 2.5888\tabularnewline
& & & & & & & Be-Nb(3) & 0.10 & 2.6360\tabularnewline
& & & & & & & Nb-Nb & 0.33 & 2.9283\tabularnewline
\hline
\end{tabular}

\end{table*}

\begin{table*}[ht]
\doublespacing
\large
\centering 

\caption{\large The computed density \emph{\(\rho\)} (in gm/cm\textsuperscript{3}),
transverse (\emph{V\textsubscript{t}} ), longitudinal
(\emph{V\textsubscript{l}} ), and average sound velocity
\emph{V\textsubscript{m}} (m/s) and Debye temperature
\emph{\(\Theta\)\textsubscript{D}} (K) of Nb\textsubscript{3}Be and
Nb\textsubscript{3}Be\textsubscript{2} phases.}
\label{table 5}
\bigskip

\begin{tabular}[]{@{}cccccc@{}}
\hline
Compounds & \emph{\(\rho\)} & \emph{V\textsubscript{t}} &
\emph{V\textsubscript{l}} & \emph{V\textsubscript{m}} &
\emph{\(\Theta\)\textsubscript{D}}\tabularnewline
\hline
Nb\textsubscript{3}Be & 7.33 & 2793.23 & 5576.25 & 3133.15 &
368.36\tabularnewline
Nb\textsubscript{3}Be\textsubscript{2} & 6.87 & 3295.04 & 6086.74 &
3677.12 & 450.91\tabularnewline
Nb\textsubscript{3}Ga & 8.33\textsuperscript{a} & 2376\textsuperscript{a} & 5219\textsuperscript{a} & 2678\textsuperscript{a} &
308.0\textsuperscript{a}, 280.71\textsuperscript{b}\tabularnewline
\hline
\end{tabular}
\\\textsuperscript{a}Ref. 45; \textsuperscript{b}Ref. 46
\end{table*}